\documentclass[twocolumn,showpacs,aps,prl,letter,amsmath,amssymb,superscriptaddress]{revtex4-1}
\usepackage{bm,color,bbm}
\usepackage{hyperref, mathtools,graphicx,natbib}
\usepackage{graphicx}
\usepackage{amsfonts}    
\usepackage{graphicx}   
\usepackage{verbatim}   
\usepackage{color}      
\usepackage{subfigure}  
\usepackage{hyperref}   
\usepackage{natbib}
\usepackage{booktabs}
\usepackage{threeparttable}
\usepackage{dcolumn}
\usepackage{multirow}
\usepackage{array}
\usepackage{bm,color,bbm}
\usepackage{hyperref, mathtools,graphicx,natbib}
\usepackage{graphicx}
\usepackage{amsfonts}    
\usepackage{graphicx}   
\usepackage{verbatim}   
\usepackage{color}      
\usepackage{subfigure}  
\usepackage{hyperref}   
\usepackage{natbib}
\usepackage{booktabs}
\usepackage{threeparttable}
\usepackage{dcolumn}
\usepackage{multirow}
\usepackage{array}
\usepackage{ulem}

\newcommand{\beq}{\begin{equation}}
\newcommand{\eeq}{\end{equation}}
\newcommand{\bqa}{\begin{eqnarray}}
\newcommand{\eqa}{\end{eqnarray}}

\definecolor{maroon}{rgb}{0.7,0,0}

\definecolor{ngreen}{rgb}{0.3,0.7,0.3}

\definecolor{golden}{rgb}{0.8,0.6,0.1}

\graphicspath{ {./figures/} }

\begin{document}
\title{Einstein-Podolsky-Rosen Steering in Two-sided Sequential Measurements with One Entangled Pair}
\author{Jie Zhu}
\affiliation{Key Laboratory of Quantum Information, University of Science and Technology of China, CAS, Hefei, 230026, China}
\affiliation{CAS Center for Excellence in Quantum Information and Quantum Physics, Hefei, 230026, China}
\author{Meng-Jun Hu}
\email{humengjun501@163.com}
\affiliation{Beijing Academy of Quantum Information Sciences, Beijing, 100089, China}
\affiliation{Key Laboratory of Quantum Information, University of Science and Technology of China, CAS, Hefei, 230026, China}
\affiliation{CAS Center for Excellence in Quantum Information and Quantum Physics, Hefei, 230026, China}

\author{Chuan-Feng Li}
\affiliation{Key Laboratory of Quantum Information, University of Science and Technology of China, CAS, Hefei, 230026, China}
\affiliation{CAS Center for Excellence in Quantum Information and Quantum Physics, Hefei, 230026, China}

\author{Guang-Can Guo}
\affiliation{Key Laboratory of Quantum Information, University of Science and Technology of China, CAS, Hefei, 230026, China}
\affiliation{CAS Center for Excellence in Quantum Information and Quantum Physics, Hefei, 230026, China}

\author{Yong-Sheng Zhang}
\email{yshzhang@ustc.edu.cn}
\affiliation{Key Laboratory of Quantum Information, University of Science and Technology of China, CAS, Hefei, 230026, China}
\affiliation{CAS Center for Excellence in Quantum Information and Quantum Physics, Hefei, 230026, China}

\begin{abstract}
Non-locality and quantum measurement are two fundamental topics in quantum theory and their interplay attracts intensive focuses since the discovery of Bell theorem. Recently, non-locality sharing among multiple observers with one entangled pair has been predicted and experimentally observed by generalized quantum measurement -- weak measurement. However, only one-sided sequential case, i.e., one Alice and multiple Bobs is widely discussed and little is known about the two-sided case. Here, we theoretically and experimentally explore the non-locality sharing in two-sided sequential measurements case in which one entangled pair is distributed to multiple Alices and Bobs. We experimentally observed double EPR steering among four observers in a photonic system. In the case that all observers adopt the same measurement strength of the weak measurement, it is observed that double EPR steering can be demonstrated simultaneously. The results not only deepen our understanding of relation between sequential measurements and non-locality but also may find important applications in many quantum information tasks, such as randomness certification.
\end{abstract}

\date{\today}

\maketitle

{\it{Introduction---}} Non-locality, which is the core characteristic of quantum theory \cite{EPR1935}, plays a fundamental role in many quantum information tasks. Bell non-locality \cite{bell,bellrmp} and Einstein-Podolsky-Rosen (EPR) steering \cite{epr,steeringrmp} are two extensively investigated notions that capture the quantum non-locality in which EPR steering is proved to be a strict general form than Bell non-locality \cite{jones2007}. From the perspective of quantum information, both Bell nonlocality and EPR steering can be demonstrated by considering two separated observers, Alice and Bob, that perform local measurements on a shared quantum state $\rho_{AB}$ and quantum non-locality is witnessed via violation of corresponding inequalities. Recently, R. Silva {\it et al} extended Bell test to include one Alice and many Bobs with intermediate Bobs performing sequential weak measurements and showed that Bell non-locality can be shared among multiple observers with one entangled pair \cite{silva2015}. Double Bell-Clauser–Horne–Shimony–Holt (CHSH) \cite{CHSH} inequality violations among three observers are then experimentally observed with one entangled photon pair by two independent groups including ours \cite{mengjunhu2018, matteo2017}. 
Based on this sequential scenario lots of works have been reported \cite{work1, work2, work3, work4, work5, work6, work7, work8, work9, work10, work11, work12} and EPR steering among multiple observers is experimentally demonstrated very recently \cite{choi2020}.

\begin{figure}[!t]
    \centering
    \includegraphics[scale=0.45]{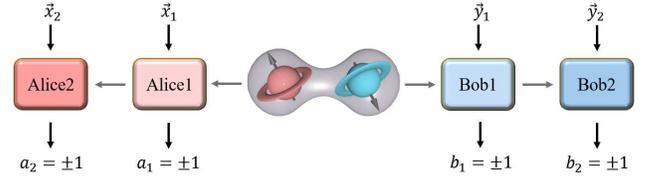}
    \caption{\textbf{Theoretical sketch.} The two-sided sequential scenario in which one entangled pair is distributed to multiple Alices and Bobs. Here $\vec{x}_{i}, \vec{y}_{j}$ represent measurement inputs and $a_{i}, b_{j}$ are corresponding dichotomic measurement outcomes, respectively. }
    \label{theory}
\end{figure}

\begin{figure*}[!t]
    \centering
    \includegraphics[scale = 0.8]{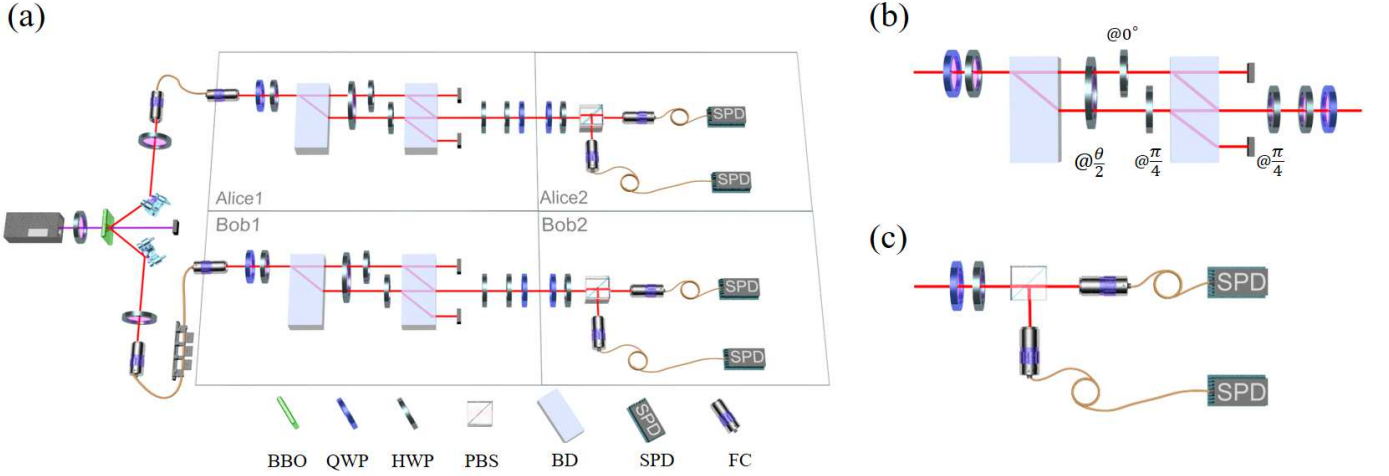}
    \caption{\textbf{Experimental setup.} {\bf (a)} Polarization-entangled photon pair are generated via the type I phase-matching spontaneous parametric down-conversion (SPDC) process by pumping a joint $\beta$-barium-borate (BBO) crystal with a 404 $\mathrm{nm}$ semiconductor laser. Signal and idler photons are then distributed to Alices and Bobs with Alice1, Bob1 performing optimal weak measurements and Alice2, Bob2 performing projective measurements.
    {\bf (b)} Setup for realizing optimal weak measurement. {\bf (c)} Setup for realizing projective measurement.
    BBO: $\beta$-barium-borate, QWP: quarter wave plate, HWP: half wave plate, PBS: polarization beam splitter, BD: beam displacer,  SPD: single photon detector, FC: fiber coupler.}
    \label{setup}
\end{figure*}

To date, however, almost all discussions are limited to one-sided sequential case, i.e., one entangled pair is distributed to one Alice and multiple Bobs. As emphasized by R. Silva {\it et al} in their last sentence in Ref. \cite{silva2015} that it would be interesting to investigate the two-sided sequential case, i.e., including multiple Alices in the setup. In this Letter, we theoretically and experimentally explore the two-sided sequential case that one entangled pair is distributed to multiple Alices and Bobs, in which middle Alices and Bobs perform optimal weak measurements and the last Alice and Bob perform projective measurement. The relation between sequential weak measurements and Bell non-locality is explicitly derived under the unbiased input condition for two-sided sequential case. It is shown that no more than two Bell-CHSH inequality violations can be obtained in the same method (see Supplementary Materials). However, here the analytical forms of EPR steering is obtained for two Alices and two Bobs case, showing that Alice1-Bob1, Alice2-Bob2 can demonstrate EPR steering simultaneously.  Using entangled photon pair, we experimentally observed double EPR steering simultaneously with $n=6$ and $n=10$ measurement settings in the case of two pairs of observers.

{\it{Theoretical framework--- }}Consider a two-party state $\rho_{AB}$ distributed to Alices and Bobs who perform sequential weak measurements as shown in Fig. 1. For convenience  of calculations and experimental realization, we choose $\rho_{AB}=|\Psi^{-}\rangle_{AB}\langle\Psi^{-}|$ with the singlet state $|\Psi^{-}\rangle_{AB}=(|\uparrow\rangle_{A}|\downarrow\rangle_{B}-|\downarrow\rangle_{A}|\uparrow\rangle_{B})/\sqrt{2}$ and require that optimal weak measurements are performed. Weak measurements can be described mathematically via positive-operator valued measure (POVM) \cite{nielsenbook} of which the Kraus operators of unbiased measurement can be settled as 
\begin{equation}
\hat{M}_{\pm 1|\Vec{k}} =\mathrm{cos}\theta|k^{\pm}\rangle\langle k^{\pm}| + \mathrm{sin}\theta|k^{\mp}\rangle\langle k^{\mp}|
\end{equation}
with dichromatic observable $\hat{\sigma}_{\vec{k}}\equiv |k^{+}\rangle\langle k^{+}|-|k^{-}\rangle\langle k^{-}|,~\langle k^+ | k^- \rangle=0$ and parameter $\theta\in [0, \pi/4]$ determines the strength of measurement. When $\theta=0$, $\hat{M}_{\pm 1|\vec{k}}$ reduces to the projector $|k^{\pm}\rangle\langle k^{\pm}|$ corresponding to project measurement, while $\theta=\pi/4$ gives $\hat{M}_{\pm 1|\vec{k}}=\hat{I}/\sqrt{2}$ representing no measurement at all. Two quantities are of particular interest in weak measurement, which are quality factor $F$ measuring the disturbance of measurement and information gain $G$ and they satisfy the trade-off relation $F^{2} + G^{2} \leq 1$ \cite{silva2015}. Measurement of Eq. (1) gives $F=\mathrm{sin}2\theta, G=\mathrm{cos}2\theta$ and satisfy optimal weak measurement condition $F^{2} + G^{2} =1$ \cite{mengjunhu2018,silva2015,sm}.

The quantum non-locality can be witnessed via violations of corresponding inequalities. Quantities measure quantum correlation need to be calculated to see whether or not they can surpass the threshold supported by local hidden variables/states theory \cite{CHSH, eprsteering}. It is shown in the following that these quantities are deeply connected to the quality factor $F$ and information gain $G$ of weak measurements. Bell quantity $I$ and EPR steering quantity $S$ both are determined by the  two-party correlation $C_{(\vec{x}, \vec{y})}=\sum_{a, b}abP(a,b|\vec{x}, \vec{y})$.
In order to obtain a general process of calculations, we first consider the joint conditional probability distribution of four observers in two-sided sequential case, which is given as
\begin{equation}
\begin{split}
&P(a_{1},a_{2}, b_{1}, b_{2}|\vec{x}_{1}, \vec{x}_{2}, \vec{y}_{1}, \vec{y}_{2})\\
=&\mathrm{Tr}[(\hat{H}_{a_{1},a_{2}|\vec{x}_{1}, \vec{x}_{2}}\otimes\hat{H}_{b_{1},b_{2}|\vec{y}_{1},\vec{y}_{2}})\rho_{AB}],
\end{split}
\end{equation}
where $\hat{H}_{a_{1},a_{2}|\vec{x}_{1},\vec{x}_{2}}\equiv \hat{M}^{\dagger}_{a_{1}|\vec{x}_{1}}\hat{\Pi}_{a_{2}|\vec{x}_{2}}\hat{M}_{a_{1}|\vec{x}_{1}}$ 
with $\hat{\Pi}$ represents projection operator and $\hat{H}_{b_{1},b_{2}|\vec{y}_{1},\vec{y}_{2}}$ is defined in the same way. The joint conditional probability distribution of any two observers is 
\begin{equation}
\begin{split}
&P(a_{i},b_{j}|\vec{x}_{i},\vec{y}_{j})  \\
= &\sum_{a_{i^{'}}b_{j^{'}}\vec{x}_{i^{'}}\vec{y}_{j^{'}}}P(\vec{x}_{i^{'}},\vec{y}_{j^{'}})P(a_{i},a_{i^{'}}, b_{j}, b_{j^{'}}|\vec{x}_{i}, \vec{x}_{i^{'}}, \vec{y}_{j}, \vec{y}_{j^{'}})
\end{split}
\end{equation}
with $i, i^{'}, j, j^{'}\in \lbrace 1, 2\rbrace$ and $i\ne i^{'}, j\ne j^{'}$. Since Alices and Bobs are independent observers $P(\vec{x}_{i},\vec{y}_{j})=P(\vec{x}_{i})P(\vec{y}_{j})$ and $P(\vec{x}_{i})=P(\vec{y}_{j})=1/n$ for unbiased inputs with $n$ is the number of measurement settings. Define correlation observable as 
\begin{equation}
\begin{split} 
&\hat{W}_{(\vec{x}_{i}, \vec{y}_{j})}  \\ =&\sum_{a_{1},a_{2},\vec{x}_{i^{'}},b_{1},b_{2},\vec{y}_{j^{'}}}a_{i}b_{j}P(\vec{x}_{i^{'}},\vec{y}_{j^{'}})\hat{H}_{a_{1},a_{2}|\vec{x}_{1},\vec{x}_{2}}\otimes\hat{H}_{b_{1},b_{2}|\vec{y}_{1},\vec{y}_{2}},
\end{split}
\end{equation}
we can obtain correlation
\begin{equation}
C_{(\vec{x}_{i}, \vec{y}_{j})} = \mathrm{Tr}[\hat{W}_{(\vec{x}_{i}, \vec{y}_{j})}\rho_{AB}].    
\end{equation}
Definition of Eq.(4) can also be used for multiple observers with the generalized definition
\begin{equation}
\begin{split}
&\hat{H}_{a_{1},...,a_{N}|\vec{x}_{1},...,\vec{x}_{N}}  \\
= &\hat{M}^{\dagger}_{a_{1}|\vec{x}_{1}}\cdots\hat{M}^{\dagger}_{a_{N-1}|\vec{x}_{N-1}}\hat{\Pi}_{a_{N}|\vec{x}_{N}}\hat{M}_{a_{N-1}|\vec{x}_{N-1}}\cdots\hat{M}_{a_{1}|\vec{x}_{1}}
\end{split}
\end{equation}
and $\hat{H}_{b_{1},...,b_{N}|\vec{y}_{1},...,\vec{y}_{N}}$ is defined in the same way as above.

The situation of EPR steering in two-sided sequential case is more complicated compared to Bell non-locality due to asymmetry of EPR steering. As a demonstration, here the calculations are limited only to two Alices and two Bobs case, in which we ask whether or not Alice1-Bob1 and Alice2-Bob2 can demonstrate EPR steering simultaneously with Alice2, Bob2 performing projective measurements.
EPR steering quantity $S$ and corresponding classical bound $B$ for $n$ measurement settings can be defined as \cite{eprsteering}
\begin{equation}
\begin{split}
&S_{n} = \frac{1}{n}|\sum_{m=1}^{n}C_{(\vec{x}^{m},\vec{y}^{m})}|, \\
&B_{n}=\mathop{\mathrm{max}}\limits_{\lbrace A_{m}\rbrace}\lbrace\lambda_{\mathrm{max}}(\frac{1}{n}\sum_{m=1}^{n}A_{m}\hat{\sigma}_{m}^{B})\rbrace,
\end{split}
\end{equation}
whereas $S_{n}>B_{n}$ refutes any local hidden states theory. Here $A_{m}\in\lbrace -1, 1\rbrace$ represents Alice's declared result for the $m-$th measurement setting of Bob's and  $\lambda_{max}(\hat{O})$ denotes the largest eigenvalue of $\hat{O}$. 
Detailed calculations give
\begin{equation}
S^{A_{1}-B_{1}}_{n_{1}} = G_{A_{1}}\cdot G_{B_{1}}.
\end{equation}
After the measurement of Alice1 and Bob1, the state becomes
\begin{equation}
\begin{split}
\rho_{\vec{k}}^{A1-B1} = \frac{1}{n_1}\sum_{\vec{k}}\sum_{i,j\in\{+,-\}}(\hat{M}_{\vec{k}}^i \otimes \hat{M}_{\vec{k}}^j)|\Psi\rangle_{AB} \langle \Psi| (\hat{M}_{\vec{k}}^i \otimes \hat{M}_{\vec{k}}^j)^\dagger,
\end{split}
\end{equation}
where the $\vec{k}$ denotes the measurement direction of Alice1 and Bob1.
Then the detailed form of the steering quantify between Alice2 and Bob2 can be obtained
\begin{equation}
\begin{split}
S^{A_{2}-B_{2}}_{n_{2}} &= 1-4(1-\frac{1}{2}F_{A_{1}} F_{B_{1}})\frac{\sum_{k,l}|\langle k^{+}|l^{+}\rangle|^{2}|\langle k^{-}|l^{+}\rangle|^{2}}{n_{1} n_{2}}  \\
-2&F_{A_{1}} F_{B_{1}} \frac{\sum_{k,l}\mathrm{Re}[\langle l^{-}|k^{+}\rangle\langle l^{+}|k^{-}\rangle\langle k^{-}|l^{-}\rangle\langle k^{+}|l^{+}\rangle]}{n_{1} n_{2}},
\end{split}
\end{equation}
where $\sum_{k,l}$ denotes double summation $\sum_{k=1}^{n_{1}}\sum_{l=1}^{n_{2}}$ with $n_{1}, n_{2}$ are numbers of measurement settings for Bob1 and Bob2 respectively, and $\{|k^{\pm}\}~(\{|l^{\pm}\rangle\})$ is the measurement basis of A1-B1 (A2-B2). Since the distributed state is the singlet state $|\psi^{-}\rangle_{AB}$, the measurement directions of Alice and Bob are chosen to be opposite to maximize the $S_{n}$. 
When measurement settings are settled $S_{n}^{A_{2}-B_{2}}$ is only determined by $F_{A_{1}}\cdot F_{B_{1}}$. Consider the case of $n=3$ in which measurement settings are chosen as $\lbrace X, Y, Z\rbrace$ and measurement strength $\theta$ is the same for Alice1 and Bob1 we can obtain that $S_{3}^{A_{1}-B_{1}}=G^{2}, S_{3}^{A_{2}-B_{2}}=1-2G^{2}/3$ with $G=\mathrm{cos}2\theta$. The corresponding classical bound is $B_{3}=1/\sqrt{3}$ and thus Alice1-Bob1, Alice2-Bob2 can both demonstrate EPR steering when $G\in (0.7598, 0.7962)$. In practice larger $n$ is needed to obtain more violations. 
It should be emphasized here that in the one-sided sequential case multiple EPR steering refers to multiple Bobs aim at steering the state of one Alice, all Bobs have to choose the same measurement settings \cite{work2,work10,choi2020}. In the two-sided sequential case, however, Bobs aim at steering the corresponding Alices and their choice of measurement settings is thus independent of each other.

\begin{figure}[!t]
    \centering
    \includegraphics[scale = 0.32]{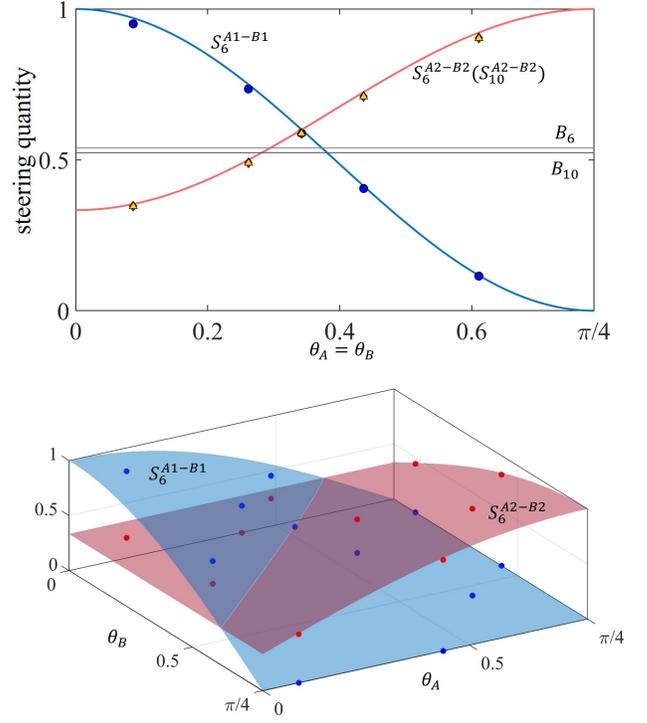}
    \caption{\textbf{Experimental results.} The steering quantities $S_6^{A_1-B1},~S_6^{A_2-B_2}$ and $S_{10}^{A_2-B_2}$ are measured with different measurement strengths of Alice1 and Bob1 ($G_{A_1} = \rm{cos}(2\theta_A)$ and $G_{B_1} = \rm{cos}(2\theta_B)$). In the upper panel, the results, that Alice1's and Bob1's measurement strengths are equal, are presented. The blue line and dots are the theoretical prediction and experimental results of $S_6^{A_1-B_1}$, respectively. The red line represents the theoretical predictions of $S_n^{A_2-B_2}$ with $n=\{6,10\}$, and the corresponding experimental results are denoted by the red rhombus and green triangle that almost overlap. The two horizontal lines are the bounds of $B_6 = 0.5393$ and $B_{10} = 0.5236$. The maximal simultaneous violation is observed when $\theta_A = \theta_B = 0.34$ and the violation values are $0.0
   493\pm0.0029$ of $S_6^{A_1-B_1}$, $0.0485\pm0.0020$ of $S_6^{A_2-B_2}$ and $0.0636\pm0.0023$ of $S_{10}^{A_2-B_2}$. The errorbars come from the Poissonian distribution of photon count that are too small to present in the figure. In the lower panel, the more experimental results with different $G_{A_1}$ and $G_{B_1}$ are presented. The blue and red surface denote the theoretical values of $S_6^{A_1-B_1}$ and $S_6^{A_2-B_2}$, and the blue and red dots are the corresponding experimental results.}
    \label{datathreed}
\end{figure}


{\it{Experimental realization---}}We now describe the experimental setup to observe non-locality sharing among four observers.  As shown in Fig. 2a, a 404 $\mathrm{nm}$ semiconductor laser with 100 $\mathrm{mW}$ power is used to pump a joint $\beta$-barium-borate (BBO) crystal to produce the polarization-entangled photon pairs via the type I phase-matching spontaneous parametric down-conversion (SPDC) process \cite{bbo}. By adjusting wave plates placed before the BBO crystal, the singlet state $|\Psi^{-}\rangle=(|H\rangle|V\rangle-|V\rangle|H\rangle)/\sqrt{2}$ is generated with $|H\rangle$ and $|V\rangle$ refer to horizontal and vertical polarization states respectively. The fidelity of the entangled pair state  is measured to be $98.76 \pm 0.08 \%$ \cite{tomography}. 
Each half of the entangled pair is coupled into different optical fibre and then distributed to Alices and Bobs with Alice1, Bob1 performing optimal weak measurements and Alice2, Bob2 performing projective measurements. Coincidence events between four detectors are registered by avalanche photodiode single-photon detectors and a coincidence counter. The joint probability distributions for different measurement settings and outcomes are extracted from these coincidence counts within 10 $\mathrm{s}$ integral time.  

As the core part of experimental setup, Fig. 2b realizes optimal weak measurements described by Kraus operators $\hat{M}_{\pm 1|\vec{k}}$ in Eq. (1). The basic idea of the setup is firstly to transform the measurement basis $\lbrace|k^{+}\rangle, |k^{-}\rangle\rbrace$ into basis $\lbrace|H\rangle,|V\rangle\rbrace$ via the basis converter consists of a quarter-wave plate (QWP) and a half-wave plate (HWP). The interference between two beam displacers (BDs) then realize optimal weak measurements $\hat{M}_{\pm 1|\vec{z}}$ with $\hat{\sigma}_{\vec{z}}\equiv|H\rangle\langle H|-|V\rangle\langle V|$ \cite{mengjunhu2018,nielsenbook}. At last another basis converter is used to transform $\{|H\rangle, |V\rangle\}$ back into the measurement basis.  To be specifically, an input state $|\phi_{k}\rangle=\alpha|k^{+}\rangle+\beta|k^{-}\rangle$ passes the basis converter placed before BD1 becomes $|\phi_{z}\rangle=\hat{R}|\phi_{k}\rangle=\alpha|H\rangle+\beta|V\rangle$. Photons with polarization $|H\rangle$ be deflected down after passing BD, while nothing happens for $|V\rangle$. BD1 can be used to entangle the path and polarization degrees of freedom of photons that $\hat{U}_{BD}|\phi_{z}\rangle=\alpha|H\rangle|d\rangle+\beta|V\rangle|u\rangle$ with $|d\rangle, |u\rangle$ represent down and up path between two BDs respectively. With operations of HWPs the state of photons before BD2 can be written as $\alpha(\mathrm{cos}\theta|V\rangle+\mathrm{sin}\theta|H\rangle)|d\rangle+\beta(\mathrm{cos}\theta|V\rangle+\mathrm{sin}\theta|H\rangle)|u\rangle$. Since only middle path out of BD2 is retained, the components $|V\rangle|d\rangle$ and $|H\rangle|u\rangle$ in state $|\varphi\rangle$ are post-selected  and path degree of freedom is eliminated. With a HWP fixed at $\pi/4$ placed after BD2 the state of photons becomes $\alpha\mathrm{cos}\theta|H\rangle+\beta\mathrm{sin}\theta|V\rangle=\hat{M}_{+1|\vec{z}}|\phi_{z}\rangle$. With another basis converter applied subsequently, the  full setup complete $\hat{M}_{+1|\vec{k}}$ operation corresponding to the $+1$ outcome of the measurement. By adjusting the HWP after BD1 from $\theta/2$ to $\pi/4-\theta/2$, operation $\hat{M}_{-1|\vec{k}}$ corresponding to the $-1$ outcome is realized \cite{mengjunhu2018}. 

We experimentally explore non-locality sharing in two-sided sequential case with two Alices and two Bobs and the results are given in Fig. 3. We firstly choose $n=6$ measurement settings in EPR steering scenario such that $B_{n=6}=0.5393$ \cite{eprsteering}. Then we consider the case that Alice1 and Bob1 measure with the $n=6$ setting but Alice2 and Bob2 measure with the $n=10$ setting. In the $(\theta_{A_{1}}, \theta_{B_{1}})$ parameter space we have chosen different points with equal strength that $\theta_{A_1}=\theta_{B_1}\in \{
\pi/36, \pi/12, 0.34, 5\pi/36, 7\pi/36\}$. Especially, the double EPR steering simultaneously can be clearly observed when $\theta_{A_{1}}=\theta_{B_{1}} = 0.34$.
The measured non-locality quantities support theoretical predictions with errors mainly come from the Poisson distribution of photon counting and imperfection of optical elements.  It is clearly shown that while Alice1-Bob1 and Alice2-Bob2 can not demonstrate Bell non-locality simultaneously (see Supplementary Materials), they can both demonstrate EPR steering with proper choice of measurement strength. 
It is interesting to point out that Alice2 and Bob2 can demonstrate one-way EPR steering if Bob1 performs no measurement and Alice1 preforms proper weak measurement \cite{yaxiao2017}.


{\it{Discussion and conclusion---}}In summary, we have explored theoretically and experimentally non-locality sharing in the two-sided sequential case with one entangled pair is distributed to multiple Alices and Bobs. We obtain the explicit formula that relates sequential optimal weak measurements and Bell quantity including one-sided sequential case as a special situation. For one-sided sequential case, it has been shown there exists measurement protocols to demonstrate arbitrary many Bell-CHSH inequality violations with biased inputs \cite{silva2015} or unequal sharpness measurement to various Bobs \cite{work12}.  It would be interesting to investigate whether or not such measurement protocols exist in two-sided sequential case. 
Due to asymmetry of Alice and Bob and freedom of choosing measurement settings, it remains an open question that whether or not there exists an elegant analytical formula for EPR steering. Specifically, it would be interesting to investigate whether or not more than two pairs of Alice-Bob can demonstrate EPR steering simultaneously in two-sided sequential case.  Using entangled photon pair, we experimentally verify the case of two Alices and two Bobs in which Alice1, Bob1 performing optimal weak measurements and Alice2, Bob2 performing projective measurements. For Alice1, Bob1 adopt the same measurement strength, we observed double EPR steering simultaneously while it is shown that double Bell-CHSH inequality violations cannot be obtained.  The results present here not only shed new light on the understanding of interplay between quantum measurement and non-locality, but also may have important applications such as unbounded randomness certification \cite{work1,coyle2019, bowles2020, new}, randomness access code \cite{das2020,giulio2020} and one-sided device independent quantum key distribution \cite{cyril2012,bennet2012, new2, new3}.


{\it{Acknowledgements---}}M.-J. Hu acknowledges H. M. Wiseman, Eric Calvencanti and Michael J. W. Hall for valuable discussions. Part of the theoretical work were done during his visiting in Griffith University. J. Zhu acknowledges X.-J. Ye, Y. Xiao, S. Cheng, and R. Wang for helpful discussions.
This work is funded by the National Natural Science Foundation of China (Grants Nos.~11674306 and 92065113) and Anhui Initiative in Quantum Information Technologies.

{\it Note:} A more genernal conclusion about CHSH-Bell inequality in sequential measurement structure has been noted in a recent work \cite{shumingcheng2020}.







{\it References ---}


\begin{thebibliography}{42}

\bibitem{EPR1935} A. Einstein, B. Podolsky, and N. Rosen, {\it Can Quantum-Mechanical Description of Physical Reality Be Considered Complete?}, Phys. Rev. \textbf{47}, 777(1935).

\bibitem{bell} J. S. Bell, {\it On the Einstein-Podolsky-Rosen paradox}, Physics, \textbf{1}, 195-200 (1935).

\bibitem{bellrmp} N. Brunner, D. Cavalcanti, S. Pironio, V. Scarani and S. Wehner, {\it Bell nonlocality}, Rev. Mod. Phys. \textbf{86}, 419 (2014).

\bibitem{epr} H. M. Wiseman, S. J. Jones, and A. C. Doherty, {\it Steering, Entanglement, Nonlocality, and the Einstein-Podolsky-Rosen Paradox}, Phys. Rev. Lett. \textbf{98}, 1404002 (2007).

\bibitem{steeringrmp} R. Uola, A. C. S. Costa, H. C. Nguyen and O. G$\ddot{u}$hne, {\it Quantum steering}, Rev. Mod. Phys. \textbf{92}, 015001 (2020).

\bibitem{jones2007} S. J. Jones, H. M. Wiseman and A. C. Doherty, {\it Entanglement, Einstein-Podolsky-Rosen correlations, Bell nonlocality, and steering}, Phys. Rev. A \textbf{76}, 052116 (2007).



\bibitem{silva2015}R. Silva, N. Gisin, Y. Guryanova and S. Popescu, {\it Multiple Observers Can Share the Nonlocality of Half of an Entangled Pair by Using Optimal Weak Measurements}, Phys. Rev. Lett. \textbf{114}, 250401 (2015).



\bibitem{CHSH} J. F. Clauser, M. A. Horne, A. Shimony, and R. A. Holt, {\it 
Proposed Experiment to Test Local Hidden-Variable Theories}, Phys. Rev. Lett. \textbf{23}, 880 (1969).

\bibitem{mengjunhu2018}M.-J. Hu, Z.-Y. Zhou, X.-M. Hu, C.-F. Li, G.-C. Guo and Y.-S. Zhang, {\it Observation of non-locality sharing among three observers with one entangled pair via optimal weak measurement}, npj Quantum Inf \textbf{4}, 63 (2018).

\bibitem{matteo2017} M. Schiavon, L. Calderaro, M. Pitaluga, G. Vallone, and P. Villoresi, {\it Three-observer
Bell inequality violation on a two-qubit entangled state}, Quantum Sci. Technol. \textbf{2}, 015010 (2017).

\bibitem{work1} F. J. Curchod, M. Johansson, R. Augusiak, M. J. Hoban, P. Wittek, and A. Acín, {\it Unbounded randomness certification using sequences of measurements}, Phys. Rev. A \textbf{95}, 020102(R) (2017).

\bibitem{work2} S. Sasmal, D. Das, S. Mal, and A. S. Majumdar, {\it Steering a single system sequentially by multiple observers}, Phys. Rev. A \textbf{98}, 012305 (2018).

\bibitem{work3} S. Datta and A. S. Majumdar, {\it Sharing of nonlocal advantage of quantum coherence by sequential observers}, Phys. Rev. A \textbf{98}, 042311 (2018).

\bibitem{work4} K. Mohan, A. Tavakoli, and N. Brunner, {\it Sequential random access codes and self-testing of quantum measurement instruments}, New J. Phys. \textbf{21}, 083034 (2019).

\bibitem{work5} H. W. Li, Y. S. Zhang, X. B. An, Z. F. Han, and G. C. Guo, {\it Three-observer classical dimension witness violation with weak measurement}, Commun. Phys. \textbf{1}, 10 (2018).

\bibitem{work6} X. B. An, H. W. Li, Z. Q. Yin, M. J. Hu, W. Huang, B. J. Xu, S. Wang, W. Chen, G. C. Guo, and Z. F. Han, {\it Experimental three-party quantum random number generator based on dimension witness violation and weak measurement}, Opt. Lett. \textbf{43}(14), 3437-3440 (2018).

\bibitem{work7} G. Foletto, L. Calderaro, A. Tavakoli, M. Schiavon, F. Picciariello, A. Cabello, P Villoresi, and G. Vallone, {\it Experimental Certification of Sustained Entanglement and Nonlocality after Sequential Measurements}, Phys. Rev. Applied, \textbf{13}, 044008 (2020).

\bibitem{work8} A. G. Maity, D. Das, A. Ghosal, A. Roy, and A. S. Majumdar, {\it Detection of genuine tripartite entanglement by multiple sequential observers}, Phys. Rev. A \textbf{101}, 042340 (2020).

\bibitem{work9} A. Kumari, and A. K. Pan, {\it Sharing nonlocality and nontrivial preparation contextuality using the same family of Bell expressions}, Phys. Rev. A \textbf{100}, 062130 (2019).

\bibitem{work10} A. Shenoy H., S. Designolle, F. Hirsch, R. Silva, N. Gisin, and N. Brunner, {\it Unbounded sequence of observers exhibiting Einstein-Podolsky-Rosen steering}, Phys. Rev. A \textbf{99}, 022317 (2019).

\bibitem{work11} C. L. Ren, T. F. Feng, D. Yao, H. F. Shi, J. L. Chen, and X. Q. Zhou, {\it Passive and active nonlocality sharing for a two-qubit system via weak measurements}, Phys. Rev. A \textbf{100}, 052121 (2019).

\bibitem{work12} P. J. Brown, and R. Colbeck, {\it Arbitrarily Many Independent Observers Can Share the Nonlocality of a Single Maximally Entangled Qubit Pair}, Phys. Rev. Lett. \textbf{125}, 090401 (2020).

\bibitem{choi2020}Y.-H. Choi, S. Hong, T. Pramanik, H.-T Lim, Y.-S. Kim, H Jung, S.-W. Han, S. Moon and Y.-M. Cho, {\it Demonstration of simultaneous quantum steering
by multiple observers via sequential weak
measurements}, Optica, \textbf{7}, 675-679 (2020). 


\bibitem{nielsenbook}M. A. Nielsen and I. L. Chuang, 2000, {\it Quantum Computation and Quantum Information} (Cambridge University, Cambridge, NY).


\bibitem{sm} see the Supplementary Materials for detailed demonstration.






\bibitem{eprsteering} D. J. Saunders, S. J. Jones, H. M. Wiseman and G. J. Pryde, {\it Experimental EPR-steering using Bell-local states}, Nat. Phys. \textbf{6}, 845-849 (2010).





\bibitem{bbo}P. G. Kwiat, E. Waks, A. G. White, I. Appelbaum, and P. H. Eberhard, {\it Ultrabright source of polarization-entangled photons}, Phys. Rev. A \textbf{60}, R773(R) (1999).
 

\bibitem{tomography}D. F. V. James, P. G. Kwiat, W. J. Munro, and A. G. White, {\it Measurement of qubits}, Phys. Rev. A \textbf{64}, 052312 (2001).



\bibitem{yaxiao2017}Y. Xiao, X.-J. Ye, K. Sun, J.-S. Xu, C.-F. Li and G.-C. Guo, {\it Demonstration of Multisetting One-Way Einstein-Podolsky-Rosen Steering in Two-Qubit Systems}, Phys. Rev. Lett \textbf{118}, 140404 (2017).
 


\bibitem{coyle2019}B. Coyle, E. Kashefi and M. J. Hoban, {\it Certified Randomness From Steering Using Sequential Measurements}, Cryptography \textbf{2019}, 3, 27.

\bibitem{bowles2020}J. Bowles, F. Baccari and A. Salavrakos, {\it Bounding sets of sequential quantum correlations and device-independent randomness certification}, Quantum \textbf{4}, 344 (2020).

\bibitem{new} G. Foletto, M. Padovan, M. Avesani, H. Tebyanian, P. Villoresi, and G. Vallone, {\it Experimental test of sequential weak measurements for certified quantum randomness extraction }, arXiv:2101.12074.

\bibitem{das2020}D. Das, A. Ghosal, S. Kanjilal, A. G. Maity and A. Roy, {\it Unbounded pairs of observers can achieve quantum advantage in random access codes with a single
pair of qubits}, arXiv: 2101.01227 [quant-ph].

\bibitem{giulio2020}G. Foletto, L. Calderaro, G. Vallone and P. Villoresi, {\it Experimental demonstration of sequential quantum random access codes}, Phys. Rev. Research \textbf{2}, 033205 (2020).


\bibitem{cyril2012}C. Branciard, E. G. Cavalcanti, S. P. Walborn, V. Scarani and H. M. Wiseman, {\it One-sided device-independent quantum key distribution: Security, feasibility, and the connection with steering}, Phys. Rev. A \textbf{85}, 010301(R) (2012).


\bibitem{bennet2012}A. J. Bennet, D. A. Evans, D. J. Saunders, C. Branciard, E. G. Cavalcanti, H. M. Wiseman, and G. J. Pryde, {\it Arbitrarily Loss-Tolerant Einstein-Podolsky-Rosen Steering Allowing a Demonstration over 1 km of Optical Fiber with No Detection Loophole}, Phys. Rev. X \textbf{2}, 031003 (2012).

\bibitem{new2} N. Walk {\it et al}, {\it Experimental demonstration of Gaussian protocols
for one-sided device-independent quantum key
distribution}, Optica, \textbf{3}, 634-642 (2016).

\bibitem{new3} T. Gehring {\it et al}, {\it Implementation of continuous-variable quantum key distribution with composable and one-sided-device-independent security against coherent attacks}, Nat. Commun. \textbf{6}, 8795 (2015).

\bibitem{shumingcheng2020} S. Cheng, L. Liu and M. J. W. Hall, {\it Limitations on sharing Bell nonlocality between sequential pairs of observers}, arXiv:2102.11574 [quant-ph].


\end{thebibliography}

\begin{thebibliography}{42}

\bibitem{silva2015}R. Silva, N. Gisin, Y. Guryanova and S. Popescu, {\it Multiple Observers Can Share the Nonlocality of Half of an Entangled Pair by Using Optimal Weak Measurements}, Phys. Rev. Lett. \textbf{114}, 250401 (2015).


\bibitem{mengjunhu2018}M.-J. Hu, Z.-Y. Zhou, X.-M. Hu, C.-F. Li, G.-C. Guo and Y.-S. Zhang, {\it Observation of non-locality sharing among three observers with one entangled pair via optimal weak measurement}, npj Quantum Inf \textbf{4}, 63 (2018).
\end{thebibliography}

\onecolumngrid
\setcounter{page}{1}
\renewcommand{\thepage}{Supplemental Material --\arabic{page}/1}
\setcounter{equation}{0}
\setcounter{figure}{0}
\setcounter{table}{3}
\renewcommand{\theequation}{S\arabic{equation}}

\section{Supplementary Materials}

\renewcommand{\thepage}{Supplemental Material --\arabic{page}/3}
\setcounter{equation}{0}
\setcounter{figure}{0}
\renewcommand{\theequation}{S\arabic{equation}}
\renewcommand{\thefigure}{S\arabic{figure}}

\section{ A. Optimal Weak Measurements}

\subsection{Weak measurement}
For a projective measurement $\{ |k^+\rangle, |k^-\rangle\}$ in a two-level system with $\langle k^+ | k^- \rangle = 0$, the projective measurement operators are $P_+ = |k^+\rangle \langle k^+|$, $P_- = |k^-\rangle \langle k^-|$ and $\sigma_k = P_+ - P_- =|k^+\rangle \langle k^+| -  |k^-\rangle \langle k^-|$. 

For a weak measurement, the positive-operator valued measurement (POVM) can be written as Kraus operators
\begin{equation}
    \hat{M}_{\pm} = {\rm{cos}}(\theta) |k^{\pm}\rangle \langle k^{\pm}| + {\rm{sin}}(\theta)|k^{\mp}\rangle \langle k^{\mp}|.
\end{equation}
For an initial state $|\psi\rangle$, the result of the weak measurement is 
\begin{equation}
\begin{split}
    \rho_{\pm} &= \hat{M}_{\pm}|\psi\rangle \langle \psi|\hat{M}_{\pm}^\dagger/{\rm{Tr}}(\hat{M}_{\pm}|\psi\rangle \langle \psi|\hat{M}_{\pm}^\dagger)\\
    &= F|\psi\rangle \langle \psi| + (1-F)(P_+|\psi\rangle \langle \psi|P_+ +P_-|\psi\rangle \langle \psi| P_-),
\end{split}
\end{equation}
with probability $P(\pm) = {\rm{Tr}}(\hat{M}_{\pm}|\psi\rangle \langle \psi|\hat{M}_{\pm}^\dagger) = G\langle \psi|P_{\pm}|\psi\rangle + \frac{1}{2}(1-G)$. The meanings of factors $F$ and $G$ are described in Ref. \cite{silva2015} and will be illustrated in the following.

\subsection{Realization of weak measurement}
For an initial two-level state $|\psi\rangle = \alpha |0\rangle + \beta |1\rangle$, a projective measurement can be realized by coupling a pointer state that can be written as 
\begin{equation}
    (\alpha |0\rangle + \beta | 1\rangle)|0\rangle_{pointer} 
    \rightarrow \alpha |0\rangle |+\rangle_{pointer} + \beta |1\rangle |-\rangle_{pointer}
\end{equation}
with $\langle + | - \rangle = 0$.
However, for a weak measurement, the evolution of the pointer state is
\begin{equation}
\begin{split}
   & |\psi\rangle = (\alpha |0\rangle + \beta |1\rangle)|0\rangle_{pointer} \\ \rightarrow & |\psi'\rangle = \alpha |0\rangle |+'\rangle_{pointer} +\beta |1\rangle |-'\rangle_{pointer},
\end{split}
\end{equation}
where $|+'\rangle = {\rm{cos}}(\theta)|0\rangle + {\rm{sin}(\theta)}|1\rangle$ and $|-'\rangle = {\rm{sin}}(\theta)|0\rangle + {\rm{cos}(\theta)}|1\rangle$, with $\langle +'|-'\rangle={\rm{sin}}(2\theta)$ indicating the strength of the weak measurement. Then a projective measurement with the basis $\{ |0\rangle, |1\rangle \}$ on the pointer state is performed. Hereto a completed weak measurement of $\hat{M}_+ = {\rm{cos}}(\theta)|0\rangle \langle 0| + {\rm{sin}}(\theta)|1\rangle \langle 1|$ and $\hat{M}_- = {\rm{cos}}(\theta)|1\rangle \langle 1| + {\rm{sin}}(\theta)|0\rangle \langle 0|$ is fullfiled. The probability of two outcomes are $P(+) = |\langle 0| \psi'\rangle|^2$ and $P(-) = |\langle 1| \psi'\rangle|^2$.

\subsection{Factors of $F$ and $G$ in the weak measurement}
In our experiment, the factors $F$ and $G$ is defined as the same as Ref. \cite{silva2015,mengjunhu2018}. $F$ denotes the disturbance of the measurement, that can be written as $F = \langle +' | -' \rangle $. $G$ denotes the information gain of the measurement, that can be written as $G = 1 - |\langle 0| -' \rangle|^2 - |\langle 1|+'\rangle|^2$, where the $|\langle 0|-'\rangle|^2$ and $|\langle 1|+'\rangle|^2$ are error rates of the weak measurement. Here the optimal condition, $F^2 +G^2 = 1$, is satisfied.

\section{ B. Bell Non-locality in Two-sided Sequential Measurements Case }

\begin{figure*}[!ht]
    \centering
    \includegraphics[scale = 0.55]{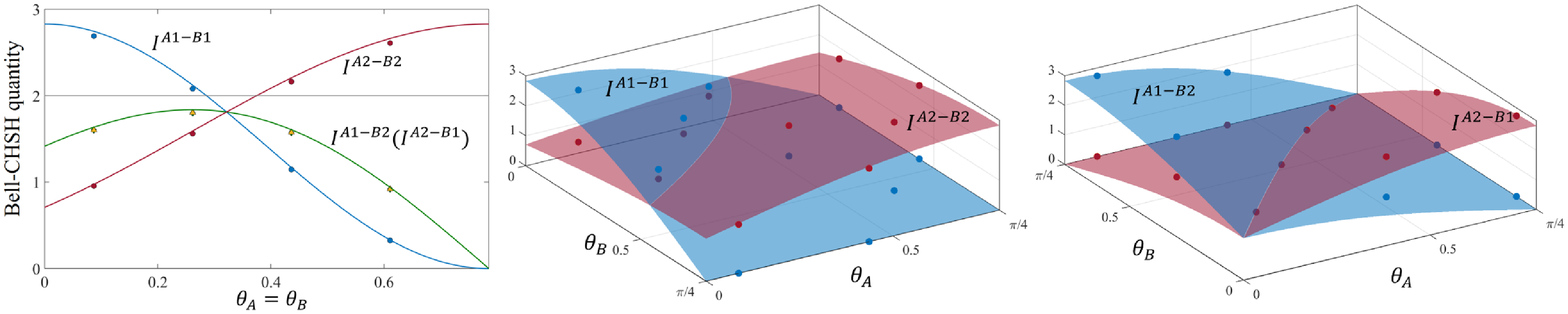}
    \caption{\textbf{Experimental results of sequential Bell test.} The results verified Eq. (S6) in which $G_{A_{2}}=G_{B_{2}}=1$ due to Alice2, Bob2 performing projective measurements. Double Bell-CHSH inequality violations is observed only when Alice1 or Bob1 performing almost no measurement such that the situation is equivalent to one-sided sequential case. When Alice1, Bob1 adopt the {\it same measurement strength}, double Bell-CHSH inequality violations can not be obtained.}
    \label{chsh}
\end{figure*}

Bell quantity $I$ in the Bell test scenario is defined as
\begin{equation}
I = |C_{(\vec{x}^{0},\vec{y}^{0})}+C_{(\vec{x}^{0},\vec{y}^{1})}+C_{(\vec{x}^{1},\vec{y}^{0})}-C_{(\vec{x}^{1},\vec{y}^{1})}|,
\end{equation}
whereas $I>2$ refutes any local hidden variables theory. Measurement settings are usually chosen as $\vec{x}^{0}=Z,~\vec{x}^{1}=X,~\vec{y}^{0}=(X-Z)/\sqrt{2},~\vec{y}^{1}=-(X+Z)/\sqrt{2}$ to reach the Tsirelson’s bound of $2\sqrt{2}$. 
In the case of two Alices and two Bobs with optimal weak measurements are performed by observers calculations based on the method of Ref. \cite{silva2015} give
\begin{equation}
\begin{split}
&I^{A_{1}-B_{1}}=2\sqrt{2}G_{A_{1}}\cdot G_{B_{1}}, \\ 
&I^{A_{2}-B_{2}}=\frac{\sqrt{2}}{2}(1+F_{A_{1}})G_{A_{2}}\cdot(1+F_{B_{1}})G_{B_{2}}, \\
&I^{A(B)_{1}-B(A)_{2}}=\sqrt{2}G_{A(B)_{1}}\cdot (1+F_{B(A)_{1}})G_{B(A)_{2}}
\end{split}
\end{equation}
Due to symmetry configuration, the Bell quantity of Alice1-Bob2 and Bob1-Alice2 have the same form.
The results are compatible with one-sided sequential case obtained by R. Silva {\it et al} if Alice1 performs no measurement with $F_{A_{1}}=1, G_{A_{1}}=0$ and Alice2, Bob2 perform projective measurements with $G_{A_{2}}=G_{B_{2}}=1$. It can be shown from the above equations that double Bell-CHSH inequality violations happens only when Alice1 or Bob1 performs almost no measurement and situation is very close to one-sided case. Furthermore, Alice1-Bob1 and Alice2-Bob2 can not demonstrate Bell non-locality simultaneously.
For the more general case with arbitrary $N$ Alices and $M$ Bobs, the explicit analytical form of Bell quantity for arbitrary Alice and Bob is derived. 
It is concluded that no more than double Bell-CHSH inequality violations can be obtained in this scenario under the unbiased input condition. 

Similarly, in the two-sided sequential case in which one entangled pair is distributed to arbitrarily many $N$ Alices and $M$ Bobs for sequential optimal weak measurements, the Bell quantity for arbitrary Alice and Bob, under the unbiased input condition, satisfies
\begin{equation}
\begin{split}
 I^{A_{r}-B_{s}}=\frac{2\sqrt{2}}{2^{(r-1)}\cdot 2^{(s-1)}}&(1+F_{A_{1}})\cdots(1+F_{A_{r-1}})\cdot G_{A_{r}}  \\
 \times&(1+F_{B_{1}})\cdots(1+F_{B_{s-1}})\cdot G_{B_{s}}
\end{split}
\end{equation}
with $r\le N, s\le M$ and $G_{A_{N}}=G_{B_{M}}=1$ if the last Alice and Bob performing projective measurements. 
For $N=1$, it naturally reduces to one-sided sequential case with one Alice and multiple Bobs.

\end{document}